# Rapid-Prototyping a Brownian Particle in an Active Bath

Jin Tae Park,[a,b] Govind Paneru,[a] Chulan Kwon,[d,*] Steve Granick,[a,c,*] and Hyuk Kyu Pak[a,b,*]

[a] Center for Soft and Living Matter, Institute for Basic Science (IBS), Ulsan 44919, South Korea
[b] Department of Physics, Ulsan National Institute of Science and Technology (UNIST), Ulsan 44919, South Korea
[c] Department of Chemistry, Ulsan National Institute of Science and Technology (UNIST), Ulsan 44919, South Korea
[d] Department of Physics, Myongji University, Yongin, Gyeonggi-Do 17058, South Korea

**Abstract**

Particles kicked by external forces to produce mobility distinct from thermal diffusion are an iconic feature of the active matter problem. Here, we map this onto a minimal model for experiment and theory covering the wide time and length scales of usual active matter systems. A particle diffusing in a harmonic potential generated by an optical trap is kicked by programmed forces with time correlation at random intervals following the Poisson process. The model's generic simplicity allows us to find conditions for which displacements are Gaussian (or not), how diffusion is perturbed (or not) by kicks, and quantifying heat dissipation to maintain the non-equilibrium steady state in an active bath. The model reproduces experimental results of tracer mobility in an active bath of swimming algal cells. It can be used as a stochastic dynamic simulator for Brownian objects in various active baths without mechanistic understanding, owing to the generic framework of the protocol

## I.     Introduction

Active particles, circulating in liquid or gas more rapidly than from ordinary Brownian motion, frequently collide with passive Brownian particles and in this way boost their mobility. Understanding this is central to the problem of swimming bacteria,[1-8] active colloidal particles,[9-11] and even catalytic enzymes.[12-14] Without wishing to minimize the important differences between these systems, they share the common feature that they involve long-range (usually hydrodynamic) interactions that are difficult or impossible either to control experimentally or compute theoretically. Complexity of the usual systems[4, 15, 16] precludes one from varying the relevant variables independently; for example, in the bacteria bath, it is not feasible to independently vary the activity of the bacteria, their concentration, and collision times with passive particles. As a result, while experimentally the probability distribution functions of particle position and displacement in bacteria systems are Gaussian in some studies and non-Gaussian in others, the relevant differences of experimental conditions are unclear[17] and analysis is problematical. Here we present a minimal model in which, experimentally and analytically, the relevant variables of the active bath are varied independently without the need for mechanistic understanding of the system. We address the strongly-overdamped situation in which momentum is a second-order effect – life at low Reynolds numbers.[18]

To model how active force modulates a thermal system,[19,20] we consider a sphere in the harmonic potential of an optical trap[21] and immersed in viscous liquid. We impose an active force produced by kicks, each of which is exerted randomly in time and decays afterwards. In experiment, the active force is generated by shifting the position of the trap center for a programmed duration time (Fig. 1a). Until another kick comes, the system reverts towards equilibrium by diffusing through the viscous bath. The simplicity of this rapid-prototyping protocol allows us to study directly the interplay between active and thermal forces, varying them independently as cannot be done in most physical systems.

We consider 1D dynamics of a tracer particle in a harmonic trap in the presence of thermal fluctuating force $\xi_{th}(t)$ and active fluctuating force $\xi_{act}(t)$ due to random kicks. In this case, the corresponding equation of motion is given by

$$\gamma \dot{x} = -kx(t) + \xi_{th}(t) + \xi_{act}(t). \tag{1}$$

Here, $k$ is the stiffness of the harmonic potential, and $\gamma$ is the dissipation coefficient representing solvent viscosity. The thermal force is white noise with zero mean and no memory. Without the active force, the system trivially has a single time constant $\tau_k = \gamma/k$, which is the equilibration time of the particle position in the harmonic potential. We impose an active force generated by a compound Poisson process and parameterized by $\tau_p$ (average time interval between kicks), $\tau_c$ (kick duration) and $X$ (root-mean-square of random kick amplitude). Considering the process with random (Poisson) arrival time, $\tau_p$ is the average Poisson time.[22,23] In the following, we generate kicks to have random amplitude with equal probability in both directions, such that the kick strength follows a Gaussian distribution with zero mean and variance $X^2$. Each kick arriving at the time $t_i$ instantaneously shifts the trap center by $x_{ci}$ which has a random amplitude $d_i$ with variance $X^2$ and decays exponentially with time constant $\tau_c$, as illustrated in Fig. 1a. Then, the position of the trap center at given time $x_c(t)$ is the sum of the past shifted positions of the center of the harmonic trap potential surviving until the next collision, given as

$$x_c(t) = \sum_i d_i e^{-(t-t_i)/\tau_c} \theta(t-t_i). \tag{2}$$

The simplicity of this rapid-prototyping protocol allows us to immediately reproduce many known features of the active matter problem. Time correlation of the active fluctuation force

$\xi_{act}(t) = kx_c(t)$ decays exponentially in steady-state for time much larger than $\tau_c$ as in other active matter systems,[1, 13] $\langle \xi_{act}(t)\xi_{act}(s)\rangle = (k^2 X^2 \tau_c/2\tau_p)e^{-|t-s|/\tau_c}$ which is derived in SI. From Eq. (2), we can show that the active force here follows the compound Poisson process

$$\dot{\xi}_{act}(t) = -\xi_{act}(t)/\tau_c + k\sum_{i=1}^{n} d_i \delta(t-t_i). \qquad (3)$$

This relation is superficially similar to Ornstein-Uhlenbeck (OU) noise, in which external noise $\xi_{ou}$ is governed by a hidden white noise $\eta_w$ and follows the stochastic equation $\dot{\xi}_{ou} = -\xi_{ou}/\tau_c + \eta_w(t)$ which yields $\langle \xi_{ou}(t)\xi_{ou}(s) \rangle \propto e^{-|t-s|/\tau_c}$.[24] Though $\xi_{act}$ and $\xi_{ou}$ give rise to the same exponential correlation, the former has higher-order cumulants for all orders, while the latter has none. The important difference is that $\eta_w$ for the OU noise is a continuous white noise, but the corresponding noise for the active force in this work is a discrete pulsed noise $k\sum_{i=1}^{n} d_i \delta(t - t_i)$. It is appealing that this simple model covers not only Poisson but also Gaussian correlated noises approaching to white noise in a particular limit, as shown in Figs. 1b-d. For the case of short force duration with $\tau_c < \tau_p$, the probability distribution function (PDF) of the active force is non-Gaussian except for short displacements around the center where it follows a Gaussian distribution (Inset of Fig. 1b). In the limit $\tau_c/\tau_p \to \infty$, the kick duration greatly exceeds the interval between kicks and the active force becomes a continuous correlated Gaussian noise due to the random nature of $d_i$ with zero mean and the OU noise is recovered (Fig. 1c).[25] We can show that the kurtosis $\propto \tau_p/\tau_c$ vanishes in this limit, which is a signal of the OU noise. A rigorous proof for the transition to the OU noise is given in Eqs. (S6)-(S9) in SI. White noise is also recovered strictly for very short force duration times (the limit $\tau_c \to 0$) and the condition of $\tau_p \to \tau_c^2$ (Fig. 1d).[26]

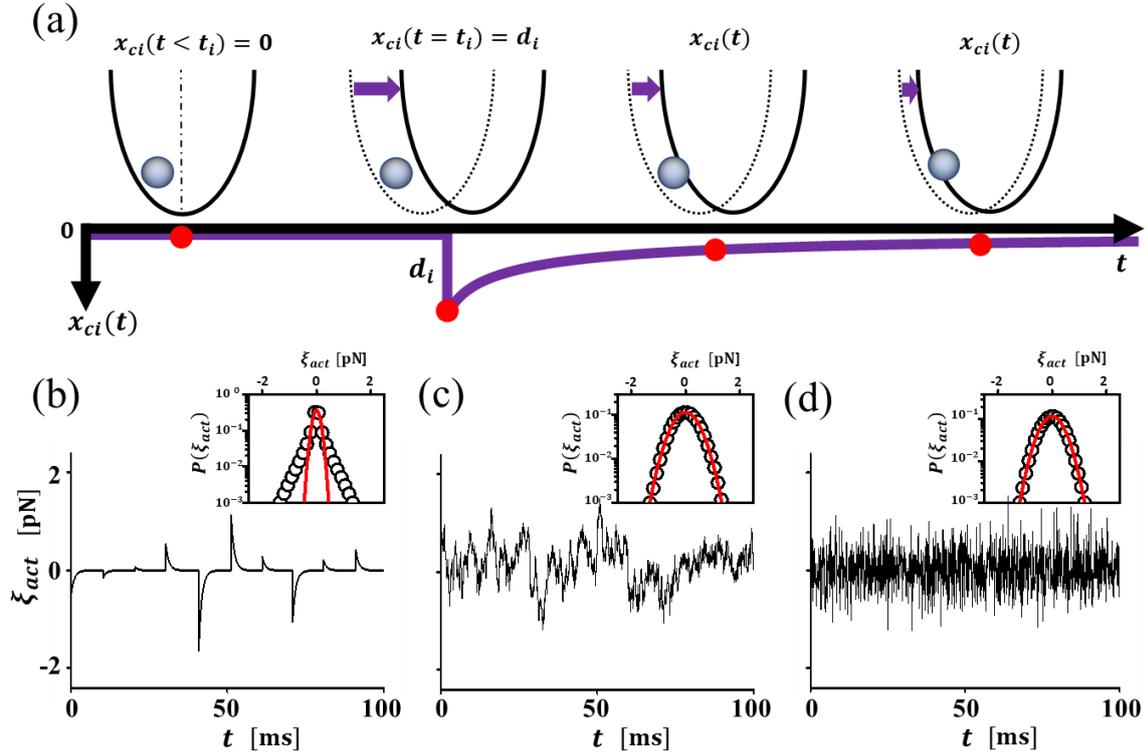

**Fig 1.** Generation of active forces. (a) Time-dependent trap center position $x_{ci}(t)$ for i[th] kick event. Temporal properties of the active force are controlled by the parameters $\tau_p$ and $\tau_c$. Here, at $t = t_i$, the trap center position is shifted from 0 to random kick position $d_i$ with variance $X^2$ and the particle subsequently relaxes towards equilibrium by diffusion through the viscous liquid. Here, $x_{ci}(t<t_i) = 0$ is chosen for the convenience of explaining the effect only from a single kick and its relaxation. But in real experiment usually $x_{ci}(t<t_i) \neq 0$ due to the previous history of kicks with relaxations. Three extreme cases are (b) Poisson correlated active forces when $0 < \tau_c < \tau_p$, (c) Gaussian correlated active forces when $\tau_c \gg \tau_p$, and (d) white noise in the limit of $\tau_c \to 0$ and $\tau_c \to \sqrt{\tau_p}$. Insets are schematic sketches of the probability distribution functions of active force $\xi_{act}$ and Gaussian fits to those curves (red solid curves).

## II. Materials and Methods

Figure 2 shows the basic scheme of the experimental setup (more details in SI).[27, 28] The harmonic potential which the tracer particle feels is created by using a computer-controlled optical tweezers, in which a 1064 nm laser is used for trapping the particle. To generate the time-dependent active force, the laser beam is fed through an acoustic optical deflector (AOD) which is controlled dynamically via a computer. The AOD is properly mounted at the back focal plane of the objective lens (100x, Oil, NA:1.4) so that the stiffness of the potential $k$ is essentially constant while shifting the position of the trap center at the speed of 5 $k$Hz. A second laser with 980 nm wavelength is used for tracking the particle position. A quadrant photodiode (QPD) is used to measure the particle position by detecting the scattered tracking laser light from the particle with the accuracy of 1 nm and at the rate of 5 $k$Hz. The sample cell consists of a highly diluted solution of $2.0\,\mu$m diameter polystyrene spheres suspended in deionized water. The volume fraction of the particles is $\phi = 2.0 \times 10^{-6}$. All experiments were carried out at $295 \pm 0.1$K. In this experiment, the trap stiffness is $k = 15.4\ p$N/$\mu$m and the characteristic relaxation time is $\tau_k = \gamma/k \cong 1.1\,m$s.

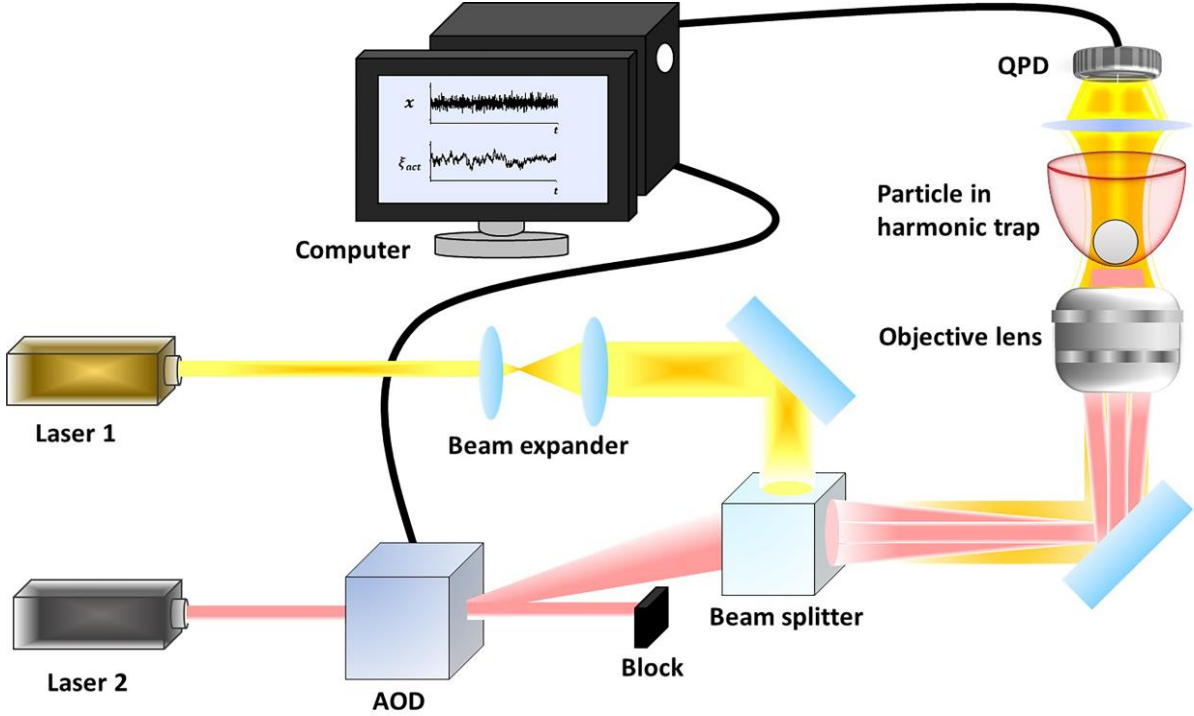

**Fig 2.** Schematic diagram of experimental set up. The Laser1 $(\lambda = 980\,n\text{m})$ and Laser2 $(\lambda = 1064\,n\text{m})$ are tracking and trapping lasers, respectively. A colloidal sphere is trapped inside the optical harmonic potential generated by the trapping beam which is manipulated by the combination of AOD (acoustic optical deflector) and computer. The QPD (quadrant photodiode) detects the position of particle and its signal is sent to computer.

## III. Results and discussion

### A. Enhanced Gaussian or non-Gaussian diffusion in an active bath.

The mean square displacement (MSD) during time interval $t$ in steady-state can be found by using the time-correlation of the active force, given as

$$<\Delta x^2(t)> = 2(k_B T/k)(1-e^{-t/\tau_k}) + 2(k_B T_{act}/k)\frac{\left[1-e^{-t/\tau_k} - (\tau_c/\tau_k)(1-e^{-t/\tau_c})\right]}{(1-\tau_c/\tau_k)}. \qquad (4)$$

Here, $T_{act} = kX^2\tau_c^2/[2k_B\tau_p\tau_k(1+\tau_c/\tau_k)]$ is "tracer activity" which quantifies how active forces modulate mobility, in a similar spirit to the phrase "active temperature" used elsewhere.[29] Trivially, MSD is governed only by the thermal diffusion coefficient $D_{th} = k_B T/\gamma$ for time much less than $\tau_c$ and $\tau_k$, given as $2D_{th}t$. In a long time limit, the MSD is $\langle\Delta x^2(t)\rangle = 2(k_B/k)\cdot(T+T_{act})$ where $D_{act} = k_B T_{act}/\gamma$ plays the formal role of active diffusion coefficient.[30] The model is formulated such that thermal and active forces split into independent contributions.[29]

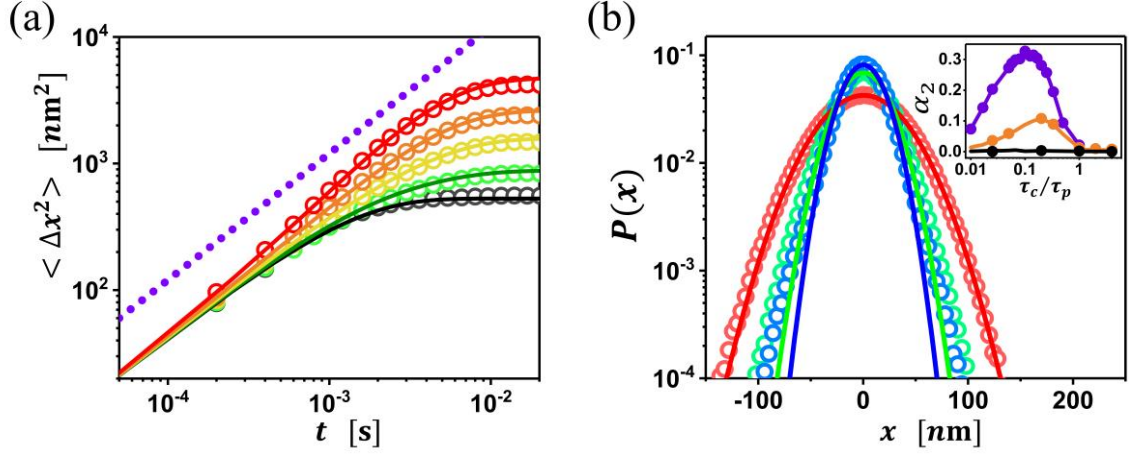

**Fig 3.** Mean square displacement (MSD) and probability distribution function (PDF) of particle in active bath. (a) MSDs are plotted against time on log-log scales, for various $\tau_c/\tau_p$: 4.0 (red), 2.0 (orange), 1.0 (yellow), and 0.33 (green). The black curve corresponds to MSD without active force. Here, $X = 37\,nm$, $\tau_c = 4\,ms$ and $\tau_k = 1.1\,ms$. The circles are experimental results and the solid curves are predictions using Eq. (4). The dotted line shows the slope of unity corresponding to free diffusion. (b) PDFs are plotted for $\tau_c/\tau_p$: 1.0 (red), 0.33 (green), and 0.2 (blue) for $X = 70\,nm$ and $\tau_c = 2\,ms$. The solid curves are Gaussian fits. The inset plots the non-Gaussian parameter $\alpha_2$ against $\tau_c/\tau_p$. The solid circles are from experiments and the solid curves are from Eq. (4). Here, $X = 70\,nm$ (violet), $37\,nm$ (orange), $20\,nm$ (black) and $\tau_c = 2\,ms$ (violet, black), $4\,ms$ (orange).

Figure 3a shows the MSD curves for various conditions. The black open circles are the MSD measurements at thermal equilibrium for $X^2 = 0$. All other colored open circles correspond to the experimental MSD data in non-equilibrium steady-state conditions with different values of $\tau_p$, but other parameter values fixed. All data agree with the theoretical predictions given in Eq. (4). As the average interval time between kicks is decreased, the number of collision events $\tau_c/\tau_p$ and the activity $T_{act}$ increase. Therefore, the saturating value of the MSD, $2\langle \Delta x^2(t) \rangle$ in Fig. 3a, increases with $T_{act}$ suggesting enhanced diffusion due to the active force.[31] The violet dotted line indicates the slope of unity for free diffusion.

The open circles in Fig. 3b are the PDFs of the particle position $P(x)$ in steady-state at fixed $\tau_c$ and $X$. For $\tau_c/\tau_p \gtrsim 1$, $P(x)$ follows a Gaussian distribution. For $\tau_c/\tau_p < 1$, $P(x)$ shows a non-Gaussian behavior. In this case, $P(x)$ has a Gaussian shape near the center but an exponential behavior at tails. The degree of non-Gaussianity of $P(x)$ can be quantified by measuring a non-Gaussian parameter,[32] defined as $\alpha_2 \equiv [<x^4>/5<x^2>^2] - 3/5$, as shown in the inset of Fig. 3b. According to this, for $\tau_c/\tau_p \gtrsim 1$, $P(x)$ is always Gaussian-like ($\alpha_2 \approx 0$). And, for $0 < \tau_c/\tau_p < 1$, $P(x)$ is non-Gaussian ($\alpha_2 > 0$). But, in the limit $\tau_c/\tau_p \approx 0$, $P(x)$ becomes Gaussian ($\alpha_2 \approx 0$) again. These results agree with previous experiments in which the PDF becomes non-Gaussian when the concentration of active particles is low, corresponding to $\tau_c/\tau_p < 1$ in our experiment.[7]

These observations can be explained by using the property of no-correlation between thermal and active forces that the particle position can be separated into non-thermal part $x_{act}$ exclusively due to the active force and thermal part $x - x_{act}$. Therefore, $P(x)$ can be expressed as the convolution of $P(x_{act})$ and $P(x - x_{act})$, $P(x) = \int P(x - x_{act}) P(x_{act}) dx_{act}$ (See Fig. S1 in SI).[33] Here, $P(x - x_{act}) = \sqrt{k/2\pi k_B T}\, e^{-k(x-x_{act})^2/2k_B T}$. The essential ingredient for non-Gaussian $P(x)$ is that $P(x_{act})$ is non-Gaussian ($\tau_c/\tau_p < 1$). However, if the tails of $P(x_{act})$ do not surpass roughly $4\sigma$ (4 standard deviations) of $P(x - x_{act})$, $P(x)$ still shows a Gaussian behavior to the eye (see Fig. S1 in SI), because the equilibration process of the particle position inside the harmonic potential wipes out the non-Gaussian effect of the active force. Even though $P(x)$ is Gaussian in this regime, the van Hove self-correlation function, $G_s(\Delta x, \Delta t)$, which is the PDF of the particle displacement $\Delta x$ during $\Delta t$, shows non-Gaussian behavior at shorter times (see Fig. S2 in SI). A recent theory explains the non-Gaussian tail by considering the hydrodynamic interactions between the tracer particle and the active particles, with the active force that arrives at random (Poisson) intervals.[34] The approach taken here complements that rigorous but complex analysis by using a rapid prototyping formalism.

### B. Comparison with real bio-active bath experiment

There were many experimental studies of real bio-active bath.[35-39] Among them, the experimental study of the tracer-particle diffusion in a quasi-two-dimensional bath of swimming algal cells (Chlamydomonas) by J. Gollub et. al.,[3] is compared with the simulation study using our model. This work is chosen because they measured MSD and van Hove correlation function of tracer particles which we are interested in for various cell concentrations.[3] The cell volume fraction $\Phi$ was controlled from $\Phi = 0.3\%$ to $7\%$ and $G_s(\Delta x, \Delta t)$ accordingly changes from highly non-Gaussian to Gaussian (colored circles in Fig. 4a). Since there is no harmonic trap ($k = 0$) in the experiment, Eq. (1) becomes $\gamma \dot{x} = \xi_{th}(t) + \xi_{act}(t)$. The corresponding MSD can be found as (see derivation in SI)

$$\left\langle \Delta x^2(t) \right\rangle = 2 D_{th} t + 2 D_{act} \left[ t - \tau_c \left( 1 - e^{-t/\tau_c} \right) \right], \qquad (5)$$

where diffusion coefficients are given as $D_{th} = k_B T / \gamma$ for thermal noise and $D_{act} = f_{RMS}^2 \tau_c^2 / (2 \gamma^2 \tau_p)$ for active force. Here, the root-mean-square (RMS) kicking force $kX$ is replaced by $f_{RMS}$ and $\left\langle \xi_{act}(t) \xi_{act}(0) \right\rangle = f_{RMS}^2 \tau_c / (2 \tau_p) e^{-t/\tau_c}$ (see derivation in SI). By fitting Eq. (5) to the MSD data of the experiment, we can estimate the values of $\tau_c = 300\,ms$ and $D_{act}$ for various values of $\Phi$. In the computer simulation using our model, $G_s(\Delta x, \Delta t)$ (solid curves) agrees well with the experimental data (colored circles) when $f_{RMS} = 0.4$ pN and $\tau_p \sim \Phi^{-3/2}$ are used (Fig. 4a). Here, $G_s(\Delta x, \Delta t)$ becomes Gaussian (non-Gaussian) when the cell concentration is high (low), corresponding to $\tau_c / \tau_p \gtrsim 1$ ($\tau_c / \tau_p < 1$). This observation agrees with the argument in the last section. When active diffusion dominates over thermal diffusion, the effective diffusion coefficient, $D_{eff} = D_{th} + D_{act}$ is approximately equal to $D_{eff} \approx D_{act} \sim \tau_p^{-1}$. Figure 4b shows an excellent agreement of $D_{eff} \sim \tau_p^{-1}$ from our simulation with $D_{eff} \sim \Phi^{3/2}$ from the experiment. The relation $\tau_p \sim \Phi^{-3/2}$ can be explained by the assumption that the average kick interval is inversely proportional to the concentration of the active particles ($\Phi^{3/2}$) in a two-dimensional active bath. The observed mean speed of the cells was $v_0 \approx 100 \,\mu\text{m/s}$ in experiment, therefore $f_{RMS}$ is roughly the magnitude of the drag force which the tracer particle feels when the speed of a tracer particle is of the same order of magnitude as that of a cell as they drag each other, $f_{RMS} \approx \gamma v_0 \approx 0.4 \text{ pN}$. This comparison suggests that even though our simple model does not include the details about the active bath, such as hydrodynamic interactions between the tracer particle and active particles[34] and among the active particles, it can explain the enhanced diffusion and non-Gaussian statistics in real bio-active bath very well.

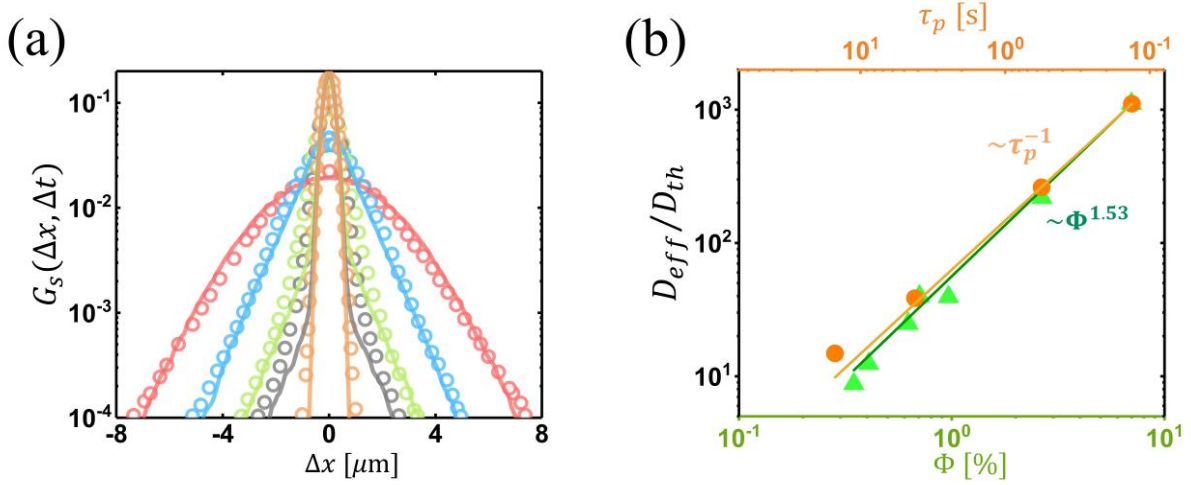

**Fig 4**. Comparison between active force model and the experimental data of algal cell active bath. (a) The van Hove self-correlation function during a fixed short time interval ($\Delta t = 0.04\,\text{s}$) at various volume fractions: $\Phi = 7.0\%$ (red), 2.7% (blue), 0.7% (green), 0.3% (gray) and $\Phi = 0\%$ (orange). The circles are from the experiment and the solid curves are from the simulation results with $\tau_p = 0.13\,\text{s}$ (red), $\tau_p = 0.56\,\text{s}$ (blue), $\tau_p = 4.2\,\text{s}$ (green) and $\tau_p = 15\,\text{s}$ (gray) at fixed $\tau_c = 0.3\,\text{s}$ and $f_{RMS} = 0.4\,p\text{N}$. (b) The effective diffusion coefficient $D_{eff}$ normalized by thermal diffusion coefficient $D_{th}$ for various volume fractions $\Phi$ in experiment and corresponding kick intervals $\tau_p$ in simulation. Here, the triangles (experiment) are fitted with $D_{eff} \sim \Phi^{1.53}$ and the circles (simulation) with $D_{eff} \sim \tau_p^{-1}$.

## C.    Heat dissipation in an active bath.

We now investigate the heat dissipation in an active bath using the model described above. Due to the energy balance between continual energy input (work) and simultaneous energy dissipation to (at) the thermal bath,[40] the first law of thermodynamics is $dE/dt = \dot{W} - \dot{Q}$. Based on the view of the potential with perturbed center by active noise, the internal energy $E$ is equal to $k(x - \xi_{act}(t)/k)^2/2$. In this situation, the rate of work done by the external agent producing active noises is known to be equal to $\dot{W} = \partial E/\partial t = -(x - \xi_{act}/k)\dot{\xi}_{act}$. Different definitions of internal energy and work can be used, but the rate of heat dissipation is uniquely defined as $\dot{Q} = (\gamma \dot{x} - \xi_{th})\dot{x}$. In non-equilibrium steady state, $\langle dE/dt \rangle = 0$, so $\langle \dot{W} \rangle = \langle \dot{Q} \rangle$, independent of the definition of heat. Theoretically, one can find $\langle \dot{Q} \rangle = k_B T_{act}/\tau_c$, which is derived in SI. To quantify this experimentally, we used the recent seminal work by Harada and Sasa,[41] in which the average rate of heat dissipation $\langle \dot{Q} \rangle$ is related to experimentally accessible quantities via a fluctuation-dissipation relation.[42-44]

$$\langle \dot{Q} \rangle = \frac{\gamma}{2\pi} \int_{-\infty}^{\infty} d\omega [C(\omega) - 2k_B T R(\omega)] \qquad (6)$$

Here, $C(\omega)$ is the velocity auto-correlation function and $R(\omega)$ is the real part of the response function for the velocity measurement in frequency space.

The integrand in Eq. (6) can be measured directly in our experiment. The noisy thin solid lines in Fig. 5a are the measurements of $C(\omega)$ and $2k_B T R(\omega)$ [45] for various conditions; the experimental detail can be found in SI. The solid thick curves are the theoretical predictions. In the equilibrium condition where $T_{act} = 0$, $C(\omega)$ and $2k_B T R(\omega)$ are superimposed satisfying the typical fluctuation-dissipation relation. As $T_{act}$ increases, however, $C(\omega)$ differs increasingly from $2k_B T R(\omega)$. $C(\omega)$ is the effect of the velocity fluctuation of the particle to increase with $T_{act}$, while the response function $R(\omega)$ is the property independent of zero mean active force, hence independent of $T_{act}$.[46]

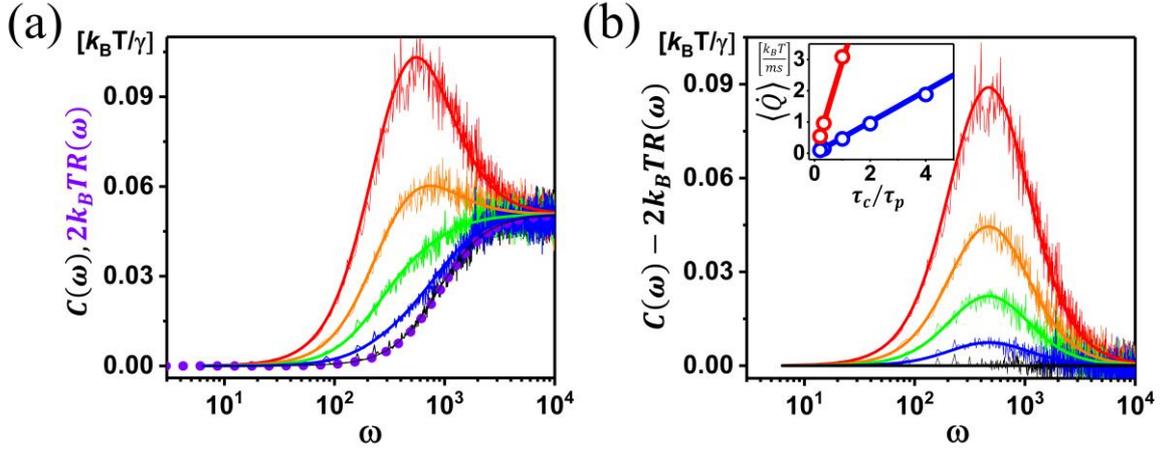

**Fig 5.** Measurement of heat dissipation rate, $\langle \dot{Q} \rangle$. (a) Correlation function $C(\omega)$ and response function $2k_B T R(\omega)$ of systems under various conditions. Here, $\tau_c/\tau_p$ : 4.0 (red), 2.0 (orange), 1.0 (green), and 0.33 (blue) with fixed values of $\tau_c = 4\,m\text{s}$ and $X = 37\,n\text{m}$. The smooth and thick solid curves are theoretical predictions for $C(\omega)$ and the noisy curves are corresponding experimental ones. The violet dashed and solid curves are the experimental and theoretical response functions. It is the same regardless of the active force. (b) Semi-log plot of spectral heat dissipation rate $C(\omega) - 2k_B T R(\omega)$ vs. frequency $\omega$ for the data in (a), in which $\tau_c$ value is fixed. Inset plots the total heat dissipation rate $\langle \dot{Q} \rangle$ against $\tau_c/\tau_p$. The open circles are experimental measurements and the solid curves are the theoretical prediction, $k_B T_{act}/\tau_c$. Here, red (blue) color corresponds to $X = 70\,n\text{m}$ ($37\,n\text{m}$) and $\tau_c = 2\,m\text{s}$ ($4\,m\text{s}$).

Figure 5b shows $C(\omega) - 2k_B T R(\omega)$ which is proportional to the spectral heat dissipation rate at given frequency $\omega$ from the data in Fig. 5a, where the correlation time $\tau_c$ is fixed but $\tau_c/\tau_p$ is varied. The area under each curve is proportional to the average value of the total heat dissipation rate $\langle \dot{Q} \rangle$ at that condition. The inset of Fig. 5b shows that the measured $\langle \dot{Q} \rangle$ agrees well with the theoretical prediction $k_B T_{act}/\tau_c$. All results related to the heat dissipation rate measurements are valid for not only Gaussian but also non-Gaussian $P(x)$. Maximum rate of heat dissipation occurs at the frequency $\omega_d = (\tau_c \tau_k)^{-1/2}$. This seems natural because $\tau_k$ is the characteristic equilibration time for the particle position to be randomized inside the harmonic potential and $\tau_c$ is the characteristic duration time of the active force which represents the duration of energy input by the active force.

## IV.   Conclusions

In summary, we demonstrated a new experimental and theoretical approach using a rapid-prototyping protocol to study systematically Brownian particles in an active bath. Our experiments using an optical trap and numerical and theoretical calculations based on the equation of motion in Eq. (1) agree in showing enhanced diffusion that gives rise to non-Boltzmann, hence non-Gaussian PDFs except for the range where the kick duration time is much larger than the average interval time between kicks. Computer simulation using our model agrees with the previously published experimental data of the tracer particle diffusion in the quasi-two-dimensional bath of swimming algal cells. We also measured heat dissipation using the fluctuation-dissipation relation and found that maximum heat dissipation rate occurs at the time-scale of the geometric mean of the kick duration time and the particle thermal equilibration time.

Our study can be used as a stochastic dynamic simulator for both the experiments and computer simulations of Brownian objects in various active baths without the need for mechanistic understanding, owing to the generic framework of this rapid prototyping protocol. This work has not considered those systems that exhibit non-Gaussian behavior at equilibrium. A vast literature considers models to describe such behavior - "strange kinetics",[47] "diffusing diffusivity",[48] continuous-time random walks,[49] phenomenology,[5, 50] and much more. In common, those situations share complexity not addressed here. First, the physical environment in those problems is not the simple harmonic potential we posit here. Second, inherent noise at equilibrium is not necessarily white noise. While these features could, in principle, be addressed using the rapid-prototype model introduced here, to do so goes beyond the scope of the present study

## Conflicts of interest

There are no conflicts to declare.

**Acknowledgements**

This work was supported by the taxpayers of South Korea through the Institute for Basic Science with grant No. IBS-R020-D1 (HKP and SG) and Basic Science Research Program of the National Research Foundation(NRF) funded by the Ministry of Education with grant No. 2020R1A2C100976111 (CK). We thank Profs. Bongsoo Kim and Jae Hyung Jeon for insightful discussion and comments.

# Supplementary Information

**Theoretical overview**

The tracer particle is kicked by a series of active forces, each of which is generated randomly at the time $t_i$ with random amplitude. Each kick shifts the center of the trap decaying exponentially with characteristic time $\tau_c$, which is schematically shown in Fig. 1. Then, the active force acting on the tracer particle at a certain time $t$ is the sum of active forces remaining until the time $t$, given as

$$\xi_{act}(t) = k\sum_i d_i e^{-(t-t_i)/\tau_c} \cdot \theta(t-t_i) = kx_c(t), \tag{S7}$$

where $\theta(t-t_i)$ is the step function equal to 1 for $t > t_i$ and 0 otherwise. $d_i$ is a Gaussian random number with variance $X^2$ and $t_i$ is produced by the Poisson distribution with mean interval $\tau_P$. The resultant position $x_c(t)$ of the center of the shifted potential is equal to the sum of the forces divided by $k$, given in the above equation.

The particle position can be expressed as $x \equiv x_{th} + x_{act}$, where $x_{th}$ is due to the thermal noise and $x_{act}$ due to the active force. Therefore, Eq. (1) in the text can be expressed as a set of two stochastic differential equations which are completely independent to each other.

$$\gamma \dot{x}_{th} = -kx_{th} + \xi_{th}, \qquad \gamma \dot{x}_{act} = -kx_{act} + \xi_{act} \tag{S8}$$

Here, $\xi_{th}(t)$ is a usual thermal noise with zero mean and time-correlation $\langle \xi_{th}(t)\xi_{th}(t') \rangle = 2\gamma k_B T \delta(t-t')$ and $\xi_{act}(t)$ is an active force in Eq. (S1). The time-correlation function of active forces is useful to investigate the stochastic properties of the system such as $\langle x^2 \rangle$, $\langle x(t)x(t') \rangle$ etc., where $\langle \cdots \rangle$ denotes the average over thermal and active forces. We find for $t' > t$

$$\langle \xi_{act}(t)\xi_{act}(t') \rangle = k^2 X^2 e^{-(t'-t)/\tau_c} \left\langle \sum_i e^{-2(t-t_i)/\tau_c} \right\rangle, \tag{S9}$$

where we use $\langle d_i d_j \rangle = X^2 \delta_{ij}$. Using the property of the Poisson distribution,

$$\left\langle \sum_i e^{-2(t-t_i)/\tau_c} \right\rangle = \int_0^t \frac{ds}{\tau_P} e^{-2(t-s)/\tau_c}. \tag{S10}$$

Using Eq. (S4) for steady-state condition ($t \gg \tau_c$), the noise auto-correlation function in Eq. (S3) takes the following form

$$\langle \xi_{act}(t)\xi_{act}(t') \rangle = k^2 X^2 \frac{\tau_c}{2\tau_P} e^{-(t'-t)/\tau_c}. \tag{S11}$$

The exponential correlation in this equation is similar to that for the OU noise, but higher-order cumulants are present for all orders, unlike the OU noise. As $\tau_c/\tau_p$ increases, the active noise becomes the OU noise, as seen in Fig. 1(c) in the text. We can observe such a tendency as the kurtosis $\propto \tau_p/\tau_c$ decreases.

One of the Referees provided a rigorous proof for the transition to the OU noise in the reviewing process. We will summarize the proof in the following. Integrating Eq. (3) from $t$ to $t+dt$, $\xi_{act}(t+dt) = \xi$ and $\xi_{act}(t) = \xi'$ are related as $\xi = \xi' - \xi'/\tau_c dt + h$, where $h = kd$ for random amplitude $d$ with probability $dt/\tau_p$ and $h = 0$ with probability $1 - dt/\tau_p$. Then, the Kolmogorov

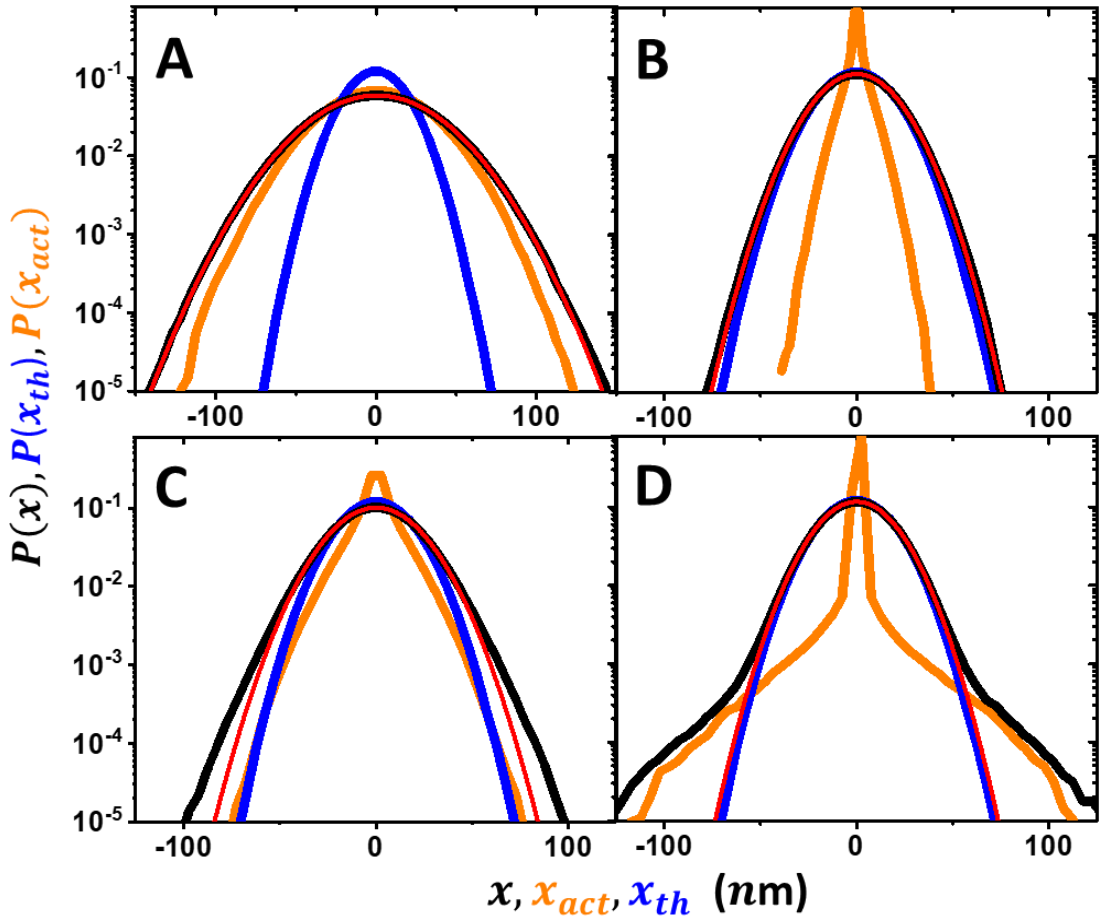

**Fig. S1. PDFs of particle position.** PDFs of $x$ (black), $x_{act}$ (orange), and $x_{th}$ (blue) from Eq. (1) and Eq. (S2) by using computer simulation. The red solid curves are Gaussian fittings to $P(x)$. (A) and (B) show the PDFs when $P(x)$ is Gaussian-like. (C) and (D) show the PDFs when $P(x)$ is non-Gaussian. The conditions are (A) $X = 20\,nm$, $\tau_c = 10\,ms$ and $\tau_p = 2\,ms$, (B) $X = 20\,nm$, $\tau_c = 2\,ms$ and $\tau_p = 10\,ms$, (C) $X = 40\,nm$, $\tau_c = 2\,ms$ and $\tau_p = 8\,ms$, and (D) $X = 70\,nm$, $\tau_c = 4\,ms$ and $\tau_p = 200\,ms$.

equation for Eq. (3) is written up to the first order in $dt$ as

$$P(\xi,t+dt) = \int d\xi' P(\xi',t)[\delta(\xi-\xi'+\xi'/\tau_c dt)(1-dt/\tau_p)$$
$$+\delta(\xi-\xi'+\xi'/\tau_c dt - kd)(dt/\tau_p)] \quad (S12)$$
$$= P(\xi,t) + \frac{dt}{\tau_p}[P(\xi-ky)-P(\xi)] + \frac{dt}{\tau_c}\frac{\partial}{\partial \xi}\xi P(\xi,t).$$

Here, $\delta(\xi-\xi'+cdt) = (1-(\partial/\partial\xi')cdt)\delta(\xi-\xi')$ is used and the integration is done over $\xi'$. Then, expanding $P(\xi-ky)$ in powers of $y$ and averaging over $y$, we get

$$\frac{\partial P(\xi,t)}{\partial t} = \frac{1}{\tau_c}\frac{\partial}{\partial \xi}\xi P(\xi,t) + \frac{1}{\tau_p}\sum_{n=1}^{\infty}\frac{k^{2n}C_{2n}}{n!}\frac{\partial^n}{\partial \xi^n}P(\xi,t). \quad (S13)$$

Here, $C_{2n} = \langle d^{2n}\rangle$ with the bracket denoting the average over $y$. By changing variables as $T = t/\tau_c$ and $X = \xi/\sqrt{\Omega}$ for $\Omega = \tau_c/\tau_p$, we find

$$\frac{\partial P(X,T)}{\partial t} = \frac{\partial}{\partial X}XP(X,T) + \sum_{n=1}^{\infty}\frac{k^{2n}C_{2n}\Omega^{1-n}}{(2n)!}\frac{\partial^{2n}}{\partial X^{2n}}P(X,T). \quad (S14)$$

In the limit $\Omega \to \infty$, we get the OU process

$$\frac{\partial P(X,T)}{\partial t} = \frac{\partial}{\partial X}XP(X,T) + \frac{k^2 C_2}{2}\frac{\partial^2}{\partial X^2}P(X,T). \quad (S15)$$

We initially prepared the system in equilibrium in the absence of the active force. Then, the PDF of $x_{th}$ remains Boltzmann every time, and hence that of $x-x_{act}(t)$ for a given $x_{act}(t)$ such that

$$P(x,t\mid x_{act}) = \sqrt{\frac{k}{2\pi k_B T}} e^{-k(x-x_{act}(t))^2/2k_B T} \quad (S16)$$

This conditional PDF of $x$ for $x_{act}(t)$ is Gaussian, while the PDF averaged over $x_{act}(t)$ becomes non-Gaussian in general.

From Eq. (S2), we have

$$x_{th}(t) = \frac{1}{\gamma}\int_0^t ds\, e^{-(t-s)/\tau_k}\xi_{th}(s), \quad x_{act}(t) = \frac{1}{\gamma}\int_0^t ds\, e^{-(t-s)/\tau_k}\xi_{act}(s). \quad (S17)$$

The variance of each position is calculated from Eqs. (S5) and (S11), given as

$$\langle x_{th}^2\rangle = \frac{k_B T}{k}, \quad \langle x_{act}^2\rangle = \frac{X^2}{2}\frac{\tau_c}{\tau_p}\left(1+\frac{\tau_k}{\tau_c}\right)^{-1} = \frac{k_B T_{act}}{k}, \quad (S18)$$

where $T_{act}$ is tracer activity which increases due to active force and the long-time limit $t \gg \tau_c$ is used. Then, we find the variance of $x = x_{th} + x_{act}$ as

$$\langle x^2 \rangle = \langle x_{th}^2 \rangle + \langle x_{act}^2 \rangle = \frac{k_B(T+T_{act})}{k}. \tag{S19}$$

The full statistics requires the information of all the moments of higher orders due to a non-Gaussian property of $x_{act}$, which is practically not possible. Fig. S1 shows the PDFs of $x$, $x_{act}$, and $x_{th}$. For large $\tau_c/\tau_p$, $P(x_{act})$ and $P(x)$ are close to Gaussian distributions because a large number of peaks in correlation time $\tau_c$ behave as Gaussian due to the central limit theorem.

### Van Hove self-correlation function $G_s(\Delta x, \Delta t)$

It is defined as the PDF of the particle displacement $\Delta x$ during $\Delta t$. We estimate the non-Gaussian parameter given in the text as $\alpha_2 \equiv [<x^4>/5<x^2>^2] - 3/5$. The result is shown in Fig. S2.

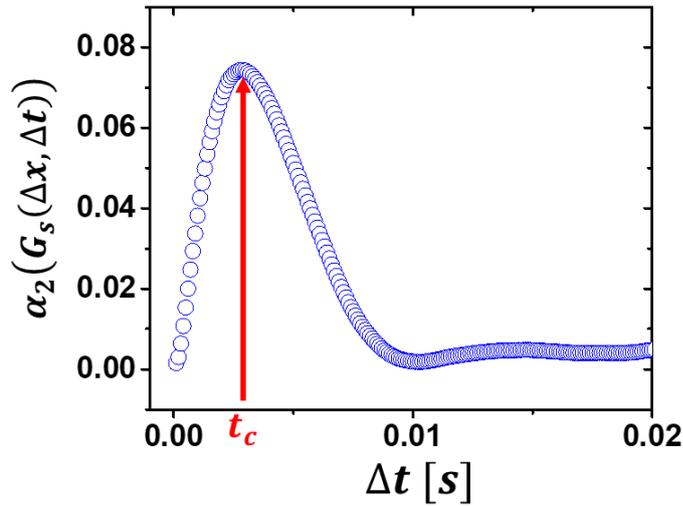

**Fig. S2.** The non-Gaussian parameter value of Van Hove self-correlation function $G_s(\Delta x, \Delta t)$ vs. $\Delta t$. $\alpha_2$ is obtained by computer simulation for $X = 70 nm$, $\tau_c = 0.3 ms$, and $\tau_k = 0.8 ms$. In the same condition, the non-Gaussian parameter value of $P(x)$ is 0.

## Mean square displacement (MSD) of a free particle in an active bath

If we consider a free particle in an active bath (without harmonic potential), the Langevin equation is described as $\gamma \dot{x}(t) = \xi_{th}(t) + \xi_{act}(t)$. Then the position of the particle $x(t)$ is expressed as

$$\Delta x(t) = \frac{1}{\gamma} \int_0^t dt' \left[ \xi_{th}(t') + \xi_{act}(t') \right], \quad (S20)$$

and the mean square displacement $\langle \Delta x^2(t) \rangle$ of the particle is

$$\langle \Delta x^2(t) \rangle = \frac{1}{\gamma^2} \int_0^t dt' \int_0^t dt'' \langle \xi_{th}(t')\xi_{th}(t'') \rangle + \frac{1}{\gamma^2} \int_0^t dt' \int_0^t dt'' \langle \xi_{act}(t')\xi_{act}(t'') \rangle. \quad (S21)$$

Using $\langle \xi_{th}(t')\xi_{th}(t'') \rangle = 2\gamma k_B T \delta(t'-t'')$, $\langle \xi_{act}(t')\xi_{act}(t'') \rangle \equiv C e^{-\left|\frac{t'-t''}{\tau_c}\right|}$, and $C = \gamma^2 \frac{D_{act}}{\tau_c} = \frac{f_{RMS}^2 \tau_c}{2\tau_p}$, we can simplify Eq. (S11). Here the $f_{RMS}$ is defined as random active kick force. (In harmonic potential, the active force satisfies this relationship; $f_{RMS} = kX$)

$$\langle \Delta x^2(t) \rangle = 2\frac{k_B T}{\gamma} t + \frac{2C\tau_c}{\gamma^2} \int_0^t dt' e^{-t'/\tau_c} \left( e^{t'/\tau_c} - 1 \right) \quad (S22)$$

Finally, the MSD of a free particle in an active bath is expressed as

$$\langle \Delta x^2(t) \rangle = 2D_{th} t + 2D_{act} \left[ t - \tau_c \left( 1 - e^{-t/\tau_c} \right) \right]. \quad (S23)$$

## Power spectral density

We can apply the Fourier transformation for the position and noise variables in a steady-state where we can neglect initial memory terms containing $e^{-t/\tau_k}$. We write $x(t) = \frac{1}{\sqrt{2\pi}} \int_{-\infty}^{\infty} d\omega e^{i\omega t} x(\omega)$ and $x(\omega) = \frac{1}{\sqrt{2\pi}} \int_{-\infty}^{\infty} d\omega e^{-i\omega t} x(t)$. Then, Eq. (S2) yields

$$x(\omega) = \frac{\xi_{th}(\omega) + \xi_{act}(\omega)}{k + i\gamma\omega}. \quad (S24)$$

Using $\langle \xi_{th}(t)\xi_{th}(t') \rangle = 2\gamma^2 D_{th} \delta(t-t')$ and $\langle \xi_{act}(t)\xi_{act}(t') \rangle = \gamma^2 D_{act} \cdot (1+\tau_c/\tau_k) \cdot e^{-|t-t'|/\tau_c}/\tau_c$ from Eq. (S3), we get

$$\langle \xi_{th}(\omega)\xi_{th}(\omega') \rangle = 2\gamma^2 D_{th} \delta(\omega + \omega'),$$
$$\langle \xi_{act}(\omega)\xi_{act}(\omega') \rangle = \frac{2\gamma^2 D_{act} \cdot (1+\tau_c/\tau_k)}{1+\tau_c^2 \omega^2} \delta(\omega + \omega'), \quad (S25)$$

where $D_{th} = k_B T/\gamma$ (diffusion constant of the medium) and $D_{act} = k_B T_{act}/\gamma$ (active diffusion coefficient). As a result, we get $\langle x(\omega)x(\omega')\rangle = S_{xx}(\omega)\delta(\omega+\omega')$ so that

$$S_{xx}(t-t') = \langle x(t)x(t')\rangle = \int_{-\infty}^{\infty} \frac{d\omega}{2\pi} C_{xx}(\omega)e^{i\omega(t-t')}. \tag{S26}$$

Using Eq. (S18), the correlation function (spectral density) of position in frequency space is given as

$$S_{xx}(\omega) = \frac{2/\tau_k}{\omega^2 + (1/\tau_k)^2}\left[\frac{k_B T}{k} + \frac{k_B T_{act}}{k}\cdot\left(1+\frac{\tau_c}{\tau_k}\right)\cdot\frac{1}{1+\omega^2\tau_c^2}\right]. \tag{S27}$$

It agrees well with the experimental results shown in Fig. S3.

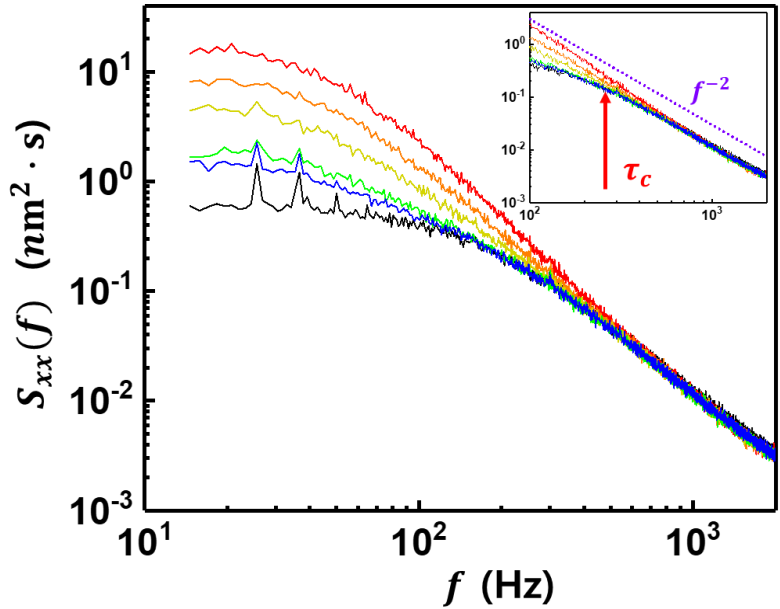

**Fig. S3. Power spectral density (PSD) of particle position for various conditions;** $\tau_c/\tau_p$ : 4.0 (red), 2.0 (orange), 1.0 (yellow), 0.33 (green), and 0.25 (blue). Black corresponds to PSD without the active force. Here, $X = 37n\text{m}$, $\tau_c = 4m\text{s}$, and $\tau_k = 1.1m\text{s}$. In the inset, the violet dotted line indicates the slope for free diffusion.

## Heat dissipation

### *The violation of fluctuation-dissipation relation*

In nonequilibrium process, incessant heat production $Q$ is found to be related to the violation of the fluctuation-dissipation theorem. Let $C(t,t') = \langle v(t)v(t') \rangle$ and $R(t,t')$ be correlation function and response functions for velocity, respectively. The violation of the fluctuation-dissipation relation (FDR) is given for $t > t'$ as

$$C(t,t') - 2k_B T R(t,t') = \frac{1}{2\gamma}[\langle v(t)(-kx(t') + \xi_{act}(t'))\rangle \\ + \langle v(t')(-kx(t) + \xi_{act}(t))\rangle]. \tag{S28}$$

The right side in this equation becomes $\gamma^{-1}\langle \dot{Q} \rangle$ in the limit $t \to t'$, where $\langle \dot{Q} \rangle = \langle \dot{x}(t)(\gamma \dot{x}(t) - \xi_{th}(t)) \rangle$ is the average rate of heat dissipation. To obtain the response function, an infinitesimal perturbation $f^p(t)$ is applied to the system. In this case, the Langevin equation is written by

$$\gamma \dot{x}(t) = -kx(t) + \xi_{th}(t) + \xi_{act}(t) + f^p(t). \tag{S29}$$

The general solution of Eq. (S23) is $x(t) = \gamma^{-1}\int_0^t dt' e^{-(t-t')/\tau_k}\left(\xi_{th}(t') + \xi_{act}(t') + f^p(t')\right)$, where the initial memory term $x(0)e^{-t/\tau_k}$ is neglected in a steady-state. The ensemble average of the velocity is $\langle v(t) \rangle = \frac{1}{\gamma}\left[ f^p(t) - \frac{1}{\tau_k}\int_0^t dt' e^{-(t-t')/\tau_k} f^p(t') \right]$ and the response function $\delta\langle v(t) \rangle / \delta f^p(t')$ is given as

$$\chi(t,t') = \frac{1}{\gamma}\left[ \delta(t-t') - \frac{1}{\tau_k} e^{-(t-t')/\tau_k} \theta(t-t') \right]. \tag{S30}$$

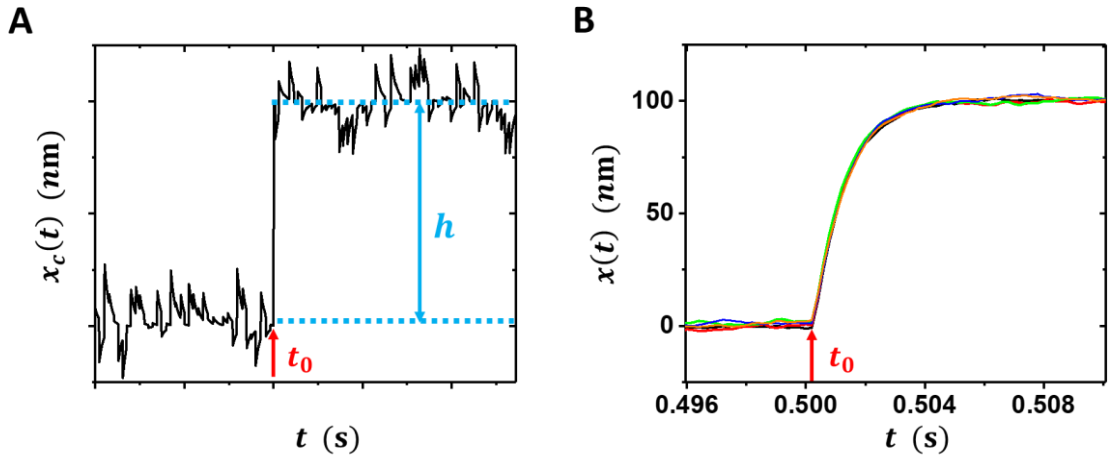

**Fig. S4. Measurement of the response function.** (A) The center of the trap is additionally shifted with distance $h$ at $t = t_0$. During this process, the system is under the active force. (B) The average trajectory of the particle over 500 different measurements. The position of the particle is observed during the relaxation process in the same condition.

Note that the response function is independent of active force as well as thermal force. Fig. S4A shows the active force and perturbation force at the time $t = t_0$. Fig. S5B shows that the presence of the active force $\xi_{act}(t)$ on the system does not change the response time $\tau_R$. In this experiment the response time is $\tau_R = \tau_k = 1.1\,ms$.

*The average rate of heat dissipation in a steady state*

The velocity correlation function in frequency space is equal to $C(\omega) = \langle v(\omega)v(-\omega) \rangle = \omega^2 S_{xx}(\omega)$. Also, the amplitude of response function can be calculated by Eq. (S23).

$$R(\omega) \equiv \mathrm{Re}(\chi(\omega)) = \frac{1}{\gamma} \cdot \frac{\omega^2}{\omega^2 + 1/\tau_k^2} \tag{S31}$$

Then, the steady-state ($t, t' \gg \tau_k$, $C(t,t') = C(t-t')$, and $R(t,t') = R(t-t')$) rate of heat dissipation is given as

$$\langle \dot{Q} \rangle = \gamma \int_{-\infty}^{\infty} \frac{d\omega}{2\pi} [C(\omega) - 2k_B T R(\omega)]. \tag{S32}$$

We write $D(\omega) \equiv \gamma(C(\omega) - 2k_B T R(\omega))$. From Eqs. (S21) and (S25), we get

$$D(\omega) = \left(\frac{2k_B T_{act}}{\tau_c^2}\right) \cdot \left(1 + \frac{\tau_c}{\tau_k}\right) \cdot \left[\frac{\omega^2}{(\omega^2 + 1/\tau_k^2)(\omega^2 + 1/\tau_c^2)}\right]. \tag{S33}$$

Then, we find the average rate of heat dissipation in a steady-state as

$$\langle \dot{Q} \rangle = \int_{-\infty}^{\infty} \frac{d\omega}{2\pi} D(\omega) = \frac{k_B T_{act}}{\tau_c}, \tag{S34}$$

where $T_{act}$ is defined in Eq. (S12). The work production rate can be defined in three different ways: (a) $\dot{W} = (\partial/\partial t) k(x - \xi_{act}(t)/k)^2/2 = -(x - \xi_{act}(t)/k)\dot{\xi}_{act}(t)$, (b) $\dot{W} = (\partial/\partial t)[kx^2/2 - x\xi_{act}(t)] = -x\dot{\xi}_{act}(t)$, and (c) $\dot{W} = \dot{x}\xi_{act}(t)$. Here, (a) is from the viewpoint of moving potential with the center at $\xi_{act}(t)/k$, (b) is defined as the summation of harmonic potential $kx^2/2$ and active potential $x\xi_{act}(t)$, and (c) is from the interpretation of $\xi_{act}(t)$ as an external time-dependent force. In all three cases, the heat dissipation rate has the same expression $\dot{x}(\gamma\dot{x} - \xi_{th}(t))$. In the steady-state limit, all definitions of work rate lead to the same average value equal to the average heat dissipation rate in Eq. (S28). The three values become different in the transient period, which is not examined in detail in our experiment but can be calculated rigorously.

*Maximum heat dissipation rate frequency*

The heat dissipation rate in frequency space has the maximum value at $\omega_d$ which we get $\partial D(\omega)/\partial \omega = 0$. We find

$$\omega_d = \frac{1}{\sqrt{\tau_k \tau_c}}. \tag{S35}$$

*Experiment setup*

Fig. S5 shows the experimental setup in detail.

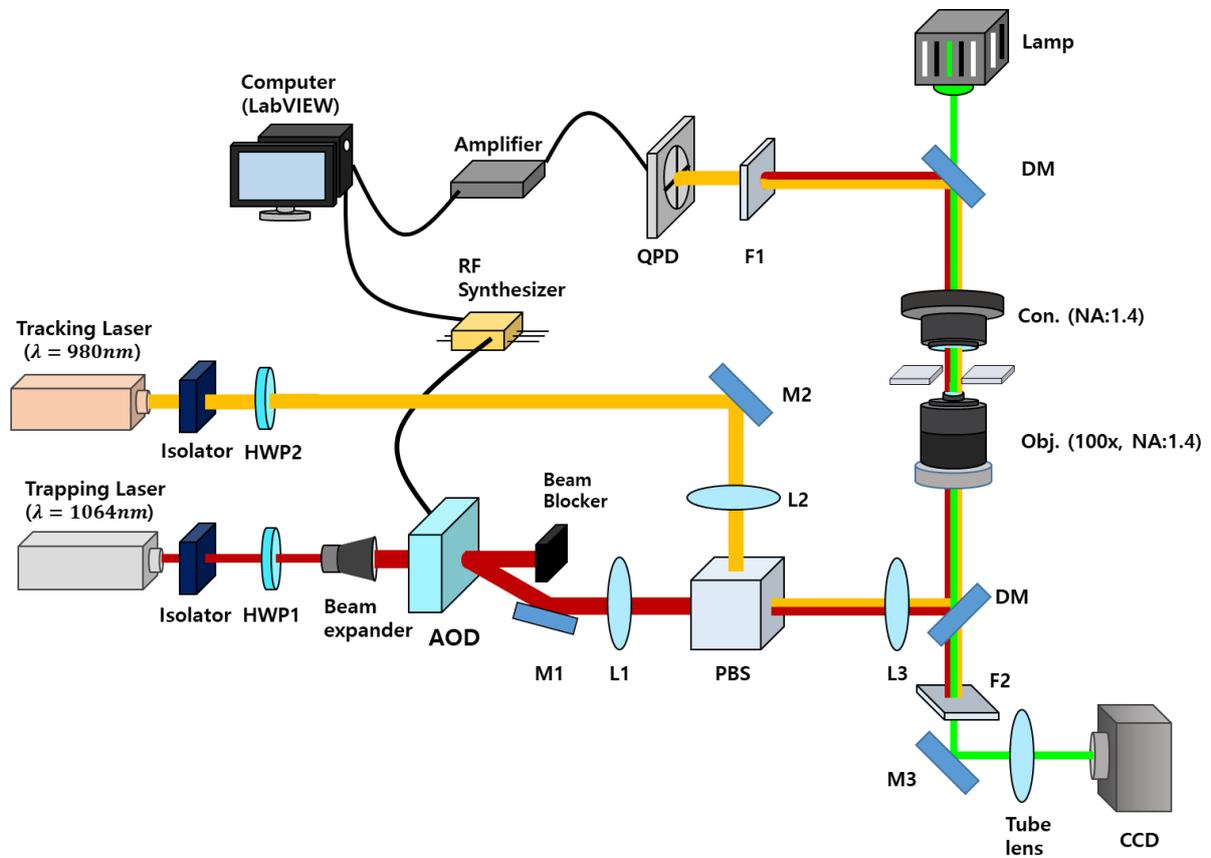

**Fig. S5. Schematic diagram of experimental setup.**